# TA-Mem: Tool-Augmented Autonomous Memory Retrieval for LLM in Long-Term Conversational QA

1st Mengwei Yuan*
*Independent Researcher*
Milpitas, USA, 95035
yuanmw1998@gmail.com

2nd Jianan Liu
*Independent Researcher*
Austin, USA, 78613
jiananliu2408@gmail.com

3rd Jing Yang
*Washington University in St. Louis*
St. Louis, USA, 63130
jing.y@wustl.edu

4th Xianyou Li
*New York University*
New York, USA, 10012
xl4230@nyu.edu

5th Weiran Yan
*Independent Researcher*
Milpitas, USA, 95035
yanwr2016@gmail.com

6th Yichao Wu
*Northeastern University*
Boston, USA, 02115
wu.yicha@northeastern.edu

7th Penghao Liang
*Northeastern University*
Boston, USA, 02115
liang.p@northeastern.edu

*Abstract*—Large Language Model (LLM) has exhibited strong reasoning ability in text-based contexts across various domains, yet the limitation of context window poses challenges for the model on long-range inference tasks and necessitates a memory storage system. While many current storage approaches have been proposed with episodic notes and graph representations of memory, retrieval methods still primarily rely on predefined workflows or static similarity top-k over embeddings. To address this inflexibility, we introduced a novel tool-augmented autonomous memory retrieval framework (TA-Mem), which contains: (1) a memory extraction LLM agent which is prompted to adaptively chuck an input into sub-context based on semantic correlation, and extract information into structured notes, (2) a multi-indexed memory database designed for different types of query methods including both key-based lookup and similarity-based retrieval, (3) a tool-augmented memory retrieval agent which explores the memory autonomously by selecting appropriate tools provided by the database based on the user input, and decides whether to proceed to the next iteration or finalizing the response after reasoning on the fetched memories. The TA-Mem is evaluated on the LoCoMo dataset, achieving significant performance improvements over existing baseline approaches. In addition, an analysis of tool use across different question types also demonstrates the adaptivity of the proposed method.

*Keywords—memory retrieval, long-context, tool-augmented, agent, memory system, large language model*

## I. INTRODUCTION

Modern large language models (LLM), built with transformer architecture, have demonstrated remarkable performance and shown great potential in natural language processing (NLP) [1]. Technologies built upon the LLM have been applied to a wide range of domains, including stock market prediction using sentiment analysis of social media posts [2] in finance, coding assistants [3] in computer science, and sentiment analysis of driver logs [4] in supply chain management, etc.

Although LLMs excel at reasoning, their effectiveness in long-term conversational QA is constrained by the context window size: smaller windows limit the available information, while larger ones can lead to irrelevant or hallucinated responses [5]. To mitigate this, existing solutions focus on creating a memory storage system and prompting the LLM only with filtered context related to the user's question [6]. Such designs consist of two components: a memory constructor and a memory retriever.

Earlier studies have primarily focused on a chunk-based memory constructor, starting with a basic memory constructor that chunks the context into smaller pieces and embeds them into vectors, where each vector is stored in parallel in the database and represents a sub-context [7]. Others improved the efficiency of this method by embedding summaries and episodic information rather than raw content [8]. Later research proposed a dynamic construction of memory in which the constructed memories are updated when new context is received [9]. Some other studies leverage the agentic design of LLM to cognitively decide the memory construction approaches [10].

In contrast to chunk-based memory systems, recent investigations find that a graph-based memory representation better resembles the human memory process by establishing direct, meaningful connections among memory pieces [11].

Despite the diverse implementations of memory constructors, retriever designs still mainly rely on vector-space similarity comparisons, which statically select the top-k relevant memory notes, or on predefined workflows that retrieve memory using a fixed traversal logic [11]. This reduces the retrieval system's flexibility and hinders the LLM's ability to adaptively access memory based on the question type. Furthermore, the predefined hyperparameter for similarity retrieval introduces unavoidable information redundancy [12], thereby reducing the cost efficiency of token use.

To address these issues, we proposed a tool augmented memory retrieval framework (TA-Mem). The system is implemented in three parts. First, we use one-shot, multi-task prompting on an agentic memory constructor to intelligently chunk the input context by detecting the topic shift, and extract memories into multiple forms in a single turn of LLM interaction. Second, a multi-indexed database supporting various query methods, including both key-based queries and vector-space similarity-based fetches, is designed to serve as a tool for the retrieval agent. Last, the retrieval agent will cognitively select and use essential tools to query the memory database based on the input question. Based on the fetched results, it determines whether additional context is needed and, if so, uses appropriate tools again, or returns the final response. This framework provides ample alternatives for the retriever to obtain the necessary context for reasoning, rather than relying on a predefined monolithic memory-fetch approach.

We evaluated our proposed framework on the LoCoMo [13] dataset. Our method achieves a significant improvement over baselines in answer quality while maintaining relatively low token usage, validating TA-Mem's effectiveness and efficiency. Furthermore, an analysis of tool usage statistics demonstrates the system's adaptability, as evidenced by a varied distribution of tool use across question domains and topics. In summary, the primary contributions of this work include:

- We address the inflexibility of LLM memory retrieval by introducing TA-Mem, a tool-augmented memory framework that enables LLM agents to autonomously explore the memory space and adaptively select tools to obtain relevant context.
- We implemented a one-shot, multi-task prompting mechanism that transforms the raw context into structured, information-dense, topic-segmented, episodic memory notes within a single LLM interaction.
- We statistically evaluated our framework on the LoCoMo dataset, demonstrating its effectiveness and efficiency.

## II. RELATED WORKS

### A. Memory System of LLM Agent

Studies on agent memories have primarily focused on storing memories into chunks or a knowledge graph. MemoryBank [8] implemented the chunk-based memory storage, inspired by the human memory process, which classifies memories into different modules based on their type. It also dynamically updates or removes the memories based on their contextual significance. During retrieval, it compares the similarity between all encoded memories and the current conversation context to select the most relevant memory. ReadAgent [7] proposed a gist memory system in which memory is chunked and summarized into short gists and prompts, and the LLM selects relevant ones for the context window, thereby enlarging context coverage during inference. MemGPT [9] implemented a memory management framework that simulates an operating system that uses the LLM as a CPU to modify and retrieve memory content. Similarly, MemoryOS [14], a recent study, focuses on controlling the memory lifecycle by storing memory in a hierarchical structure and updating it based on its temporal significance. A-Mem [10] uses agentic design for memory generation, in which the LLM processes the context into multiple notes and empowers it to dynamically decide when and which memory notes to update. Other than these chunk-based approaches, Mem0 [11] advocates graph storage of memory, in which memories are extracted into a graph, with entities, relationships, and tags extracted from conversation history representing the nodes, edges, and labels in the graph. The memory is retrieved by first locating the relevant node through semantic similarity, then traversing the outgoing and incoming nodes connected with it. Regardless of the memory storage format, the similarity top-k search has been widely adopted in retrieval.

### B. LLM Agent Tools

The LLM agent's tool integration has substantially expanded its capacity and predictability. However, simply adding available tools for an LLM agent does not scale, as LLMs tend to hallucinate about tool choices, and less effective tools will dilute the overall context and bias the LLM's decision by adding unnecessary ambiguity [15]. Studies have proposed addressing this by introducing a tool preparation step that only exposes relevant tools for the LLM to select, using a filter system or hierarchical structures of tool definition [16]. Latest works start to prompt the LLM to create its own tool through a create, decide, execute, and rectify process so that the LLM examines the tool's effectiveness by testing it through execution and improving it over iterations through rectification [17].

## III. METHODOLOGY

The overall architecture of the TA-Mem framework is shown in Figure 1, where the QA task is performed in three stages: memory extraction, storage, and retrieval.

### A. Epsodic Memory Constructor

The memory extraction agent is prompted with instructions and examples and obtains all necessary information for memory storage in a single LLM interaction. It detects topic shifts from the conversation history, segments the input into multiple chunks, and extracts episodic information for each chunk based on semantic closeness, as in (1).

$$LLM(P_e, C_n) \rightarrow \{N_1, N_2, ..., N_k\} \quad (1)$$

where $P_e$ is the extraction prompt, emphasizing the detection of the semantic boundaries within input context, and $C_n$ is the $n^{th}$ input context. Each extracted memory note $N_i$ can be represented as:

$$N_i = \{m, n, S_i, K_i, P_i, F_i, (E_i, t_i), T_i\} \quad (2)$$

where m, n are the start and end index of messages that $N_i$ is corresponding to respectively. We implemented an overlapping mechanism to smooth the chunk boundaries, so that adjacent notes have a small amount of overlap. $S_i$ is the summary of a segmented sub-context; $K_i$ is the set of semantically important keywords that appeared in the original context, $P_i$ is the set of people involved and mentioned, $F_i$ is the list of facts associated with certain persons, $E_i$ is the list of events, associated with $t_i$ which are the temporal references deducted from the context, and $T_i$ is the semantic tag generated for the sub-context. Based on the start and end message index indicated by LLM, we construct the final memory page $M_i$ as:

$$M_i = \{c_{mn}, N_i, t\} \quad (3)$$

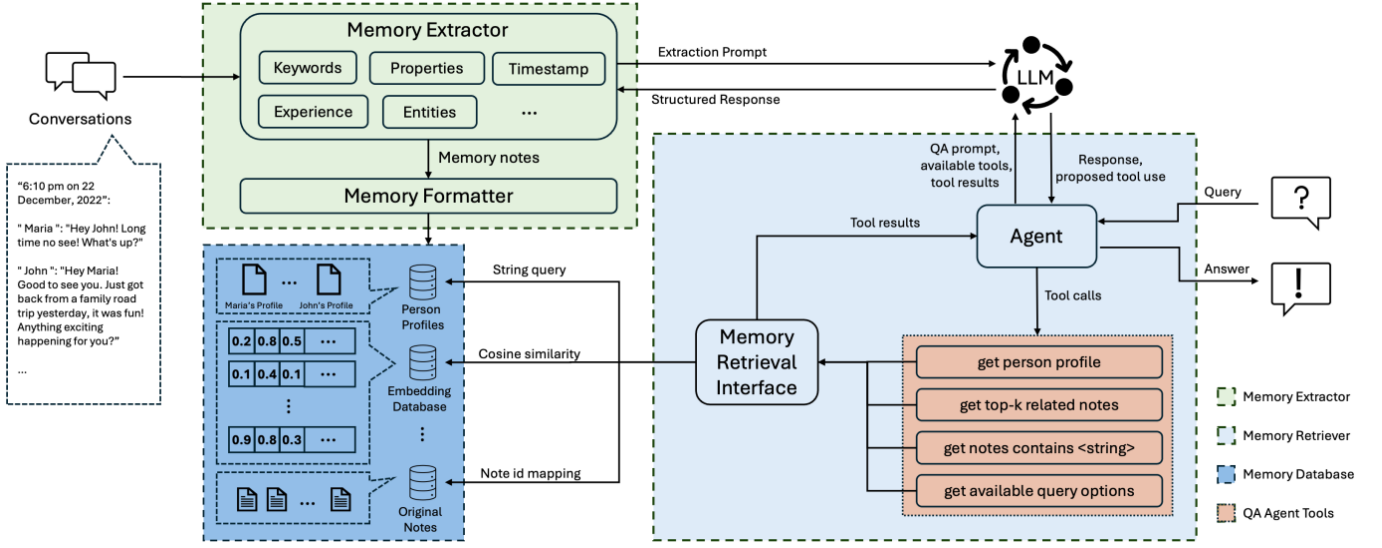

Fig. 1. Architecture of the TA-Mem framework. The memory extractor first completes offline memory processing to semantically chunk the conversation dialogues and extract episodic information. Then, a database constructor stores and indexes the memory page through different keys and embeddings. The database consists of multiple query interfaces, allowing it to integrate with the tools of a retrieval agent. The retriever reasons on user input questions by running an agentic loop with tools to autonomously explore the memory database and precisely locate the relevant memory context for model inference.

where $c_{mn}$ is the original dialogue from index m to n, and t is the conversation timestamp. This memory page representation ensures that a temporal reference is always associated with the conversation session and with each event or activity mentioned in it, which helps the QA agent resolve conflicts or outdated memory and select the latest information based on timestamps, guided by the prompt. The original messages also prevent the reasoning process from being biased by the extracted content, thereby maintaining information retention during QA.

### B. Multi-Indexed Database with Tools

To interface with the retrieval agent and diversify memory access methods, the extracted memory is stored in the database and indexed with different keys, including string representations of a person's name, semantic tags, keywords, and vector representations of events and facts. They can be queried by key matching:

$$Q_s(v, \tau) = \{M_i \mid v \in M_i, v \in V_\tau\} \quad (4)$$

where v is the key of the string query $Q_s$, and $\tau$ is the type parameter with three possible values (person, tag, keyword), indicating the key set $V_\tau$ to match against. Memory pages can also be queried through cosine similarity comparison of events or facts:

$$Q_k(v, \tau, k) = \{M_i \mid v_i \in M_i \text{ Rank}(\frac{e_v \cdot e_i}{|e_v||e_i|}) \leq k\} \quad (5)$$

where $\tau$ indicates whether the top-k based query $Q_k$ is fetching the page by comparing events or facts similarity, $v_i$ are the events or facts records in the database, and $e_i$ are their embedding representations.

In addition to memory page fetching, the database also collects all events and their timestamps associated with a person in a set $E_p$, and facts in a set $F_p$, allowing the retriever to track a person's experience or profile through:

$$Q_p(v, \tau) = r, \ r \in \{E_p, F_p\} \quad (6)$$

However, it is still a challenge for LLM to pick up the exact word or phrases during a string query due to the lexical variation under the same semantic meaning. To guide the LLM in selecting appropriate words or names in the string and person queries, we implemented a feature that provides all available keys to the retriever as a reference set, and added instructions in the retriever prompt to encourage the retrieval agent refer to the key set in the next attempt if the query returns nothing.

### C. Memory Retrieval Agent

In the retrieval module, we implemented an agentic loop to handle the question. First, the agent receives the user's query and combines it with the system prompt, which includes descriptions of all available tools and instructions for QA tasks. Then it decides which tool to use and generates the required parameters for that tool. Second, it uses selected tools to fetch context via the memory database's query interface. After reasoning on the fetched result, the LLM will have two possible decisions: continue with additional tool calls or provide the final answer.

To reduce unnecessary memory duplication in the context window and improve the efficiency of token usage, the retriever module maintains a per-QA-session memory cache that temporarily stores previously retrieved memory pages and person profiles, and drops the content from the tool result when it is found in the cache.

## IV. EXPERIMENTS

### A. Dataset and Baselines

The framework is evaluated on LoCoMo [13] dataset, which contains 10 very long-term conversations and 1986 questions in total. Each conversation involves two speakers, spans multiple sessions, covers a range of topics, and is considered an ideal evaluation of the model's long-range inference ability.

TABLE I. PERFORMANCE COMPARISON WITH EXISTING BENCHMARKS ON LOCOMO DATASET.

| Method | Multi-Hop | | Temporal | | Open Domain | | Single-Hop | | Token Usage |
|---|---|---|---|---|---|---|---|---|---|
| | F1 | BLEU-1 | F1 | BLEU-1 | F1 | BLEU-1 | F1 | BLEU-1 | |
| MemoryBank | 5.00 | 4.77 | 9.68 | 6.99 | 5.56 | 5.94 | 6.61 | 5.16 | 432 |
| ReadAgent | 9.15 | 6.48 | 12.60 | 8.87 | 5.31 | 5.12 | 9.67 | 7.66 | 643 |
| LoCoMo | 25.02 | 19.75 | 18.41 | 14.77 | 12.04 | 11.16 | 40.36 | 29.05 | 16910 |
| MemGPT | 26.65 | 17.72 | 25.52 | 19.44 | 9.15 | 7.44 | 41.04 | 34.34 | 16977 |
| Mem0 | **38.72** | 27.13 | 48.93 | 40.51 | **28.64** | 21.58 | 47.65 | 38.72 | 1764 |
| A-Mem | 27.02 | 20.09 | 45.85 | 36.67 | 12.14 | 12.00 | 44.65 | 37.06 | 2520 |
| MemoryOS | 35.27 | 25.22 | 41.15 | 30.76 | 20.02 | 16.52 | **48.62** | **42.99** | 3874 |
| TA-Mem | 35.62 | **27.84** | **55.95** | **51.47** | 26.42 | **21.82** | 44.87 | 38.74 | 3755 |

The question set comprises five categories: multi-hop, temporal, open-domain, single-hop, and adversarial. We excluded adversarial questions from quality analysis, as this category is designed to trick the model into giving incorrect answers, and the dataset provides no ground truth. We compared our model with existing benchmarks on the LoCoMo dataset, including MemoryBank [8], ReadAgent [7], LoCoMo [13], MemGPT [9], Mem0 [11], A-Mem [10], and MemoryOS [14].

*B. Setup and Evaluation Metrics*

The proposed framework is implemented with the GPT-4o-mini model for both memory extraction and retrieval. The encoding model all-MiniLM-L6-v2 is used to embed the events and facts. In the similarity top-k query, we set k to 5 in our experiments. We also limited the retrieval agent's maximum turns to 7 to avoid infinite agentic loops and hallucinatory output.

The F1 and BLEU-1 scores are used to measure the quality of the answer against the provided ground truth. To evaluate the efficiency of the TA-Mem, we also reported the average token usage for QA across the first 4 categories and the average number of turns it cost across all question types.

*C. Result and Analysis*

Experimental results are shown in Table 1. Our framework outperforms all other benchmarks and achieves a noticeable performance gain on the temporal question, with F1 scores of 55.95 and BLEU-1 scores of 51.47. The proposed framework also achieves the highest BLEU-1 score on multi-hop and open-domain questions across all benchmarks and performs competitively on single-hop questions. Despite the agentic loop design naturally increasing the number of LLM interactions, the proposed method still maintains token efficiency, with an average token count of 3755 per question. This indicates the advantage of tool use for concentrating information and filtering context compared with traditional monolithic top-k retrieval. Using appropriately defined tools, it provides a more granular, denser representation of context and enables the model to extend its reasoning capacity on the retrieval process.

Furthermore, we examined the distribution of tool use across different question types and within each question type, including the adversarial category (Figure 2). The average number of iterations to answer a question is 2.71 (excluding the adversarial type). The patterns of tool use differ significantly across types, with temporal questions usually semantically associated with activities or events, resulting in a large portion of the tool calls focusing on event queries. The open-domain question, by contrast, focuses mainly on querying for facts, where relevant memories are more likely to be stored. This distribution implies that different question types require different memory exploration strategies, as the relevant contextual information varies across aspects of the memory, and indicates the high adaptability of our memory framework.

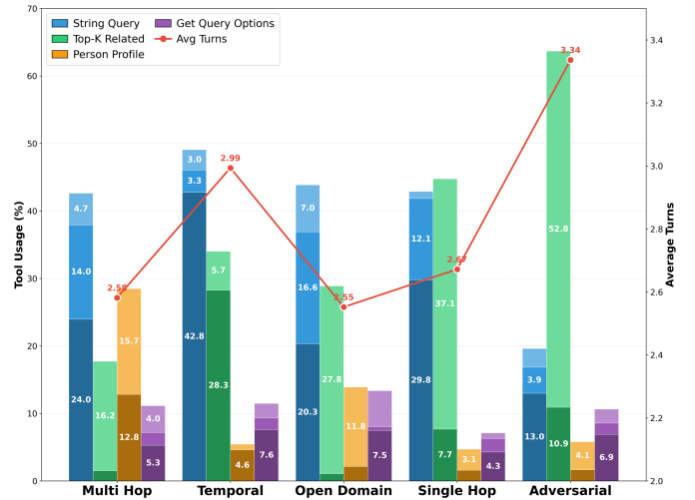

Fig. 2. Tool distributions over 5 question categories and the average number of turns. From dark to light: String query (blue) - keyword match, mentioned person match, semantic tag match; top-k retrieval (green) - events match, fact match; person's profile query (yellow) - get person's events, get person's facts; available query options (purple) - get available keywords, get available person names, get available semantic tags.

*D. Ablation Study*

We conducted an ablation study on the iteration budget to analyze the tradeoff between performance and computational cost. Figure 3 shows that performance starts to converge after the budget exceeds 4 iterations. Token usage is also capped at around 4 iterations, as 97.73% questions are finalized by then, and the memory page caching mechanism also prevents duplicate fetching. We also compared the success rate across different iteration budgets, shown in Figure 3, defined as the percentage of questions that the agent autonomously decided to finalize (no more tool calls) before reaching the budget limit. Though the major gain is observed around 4 to 5 iterations, and a higher iteration budget introduces latency for questions that require more rounds of reasoning, we set the retriever agent iteration limit to 7 to obtain the maximum performance. The result indicates the retriever agent's robustness in converging on QA reasoning and its token efficiency in a multi-iteration agent loop.

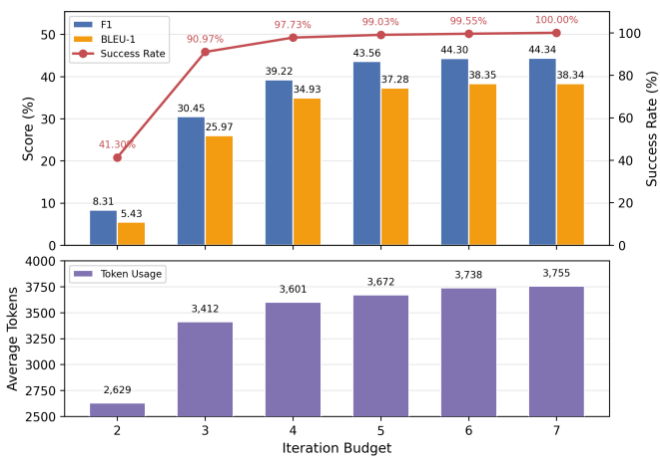

Fig. 3. Success rate, F1, BLEU-1, and token usage vs iteration budgets.

Another ablation study is conducted to compare the effectiveness of chunking methods. We compared our LLM agent-based method with fixed-length (512-token) chunking with 25% overlap and semantic chunking (split when below the 8th percentile, which results in a similar number of chunks in total compared with our method), by removing only the chunking instruction from the extractor prompt and using the chunked context as input to the extractor module. The fixed-length has an overall F1 score of 35.34%, and BLEU-1 of 29.21%; semantic chunking achieves 43.73% F1 and 38.39%, while our method achieves 44.34% F1 and 38.34% BLEU-1. This indicates that, despite potential prompt-related inconsistencies, our method not only combines chunking and memory extraction into a single step but also generates valid chunks and achieves competitive performance.

## V. Conclusion

In this paper, we propose a novel memory framework, TA-Mem, that addresses the LLM's limitation in reasoning over long-term conversational context by leveraging a sophisticated memory storage agent that provides improved memory granularity and a tool-augmented retrieval agent that cognitively retrieves relevant information to infer the answer to the user question. The framework enables the memory retrieval process to autonomously explore memory storage via a multi-indexed database, rather than relying solely on similarity comparison. The evaluation results confirm that our framework generates high-quality answers to questions requiring long-range contextual inference while maintaining high token efficiency, and the variance in the distribution of tool use also demonstrates its adaptability across different question types, indicating the feasibility of integrating the tool into memory system design for LLMs.

Although the empirical results show a performance gain for the TA-Mem framework, certain limitations remain. First, the extractor's performance depends on the prompt, introducing inconsistency and necessitating fine-tuning the instructions. Moreover, the agentic loop method for QA introduces latency to the overall system, bringing challenges for time-sensitive applications. Future work will investigate scaling this method to significantly larger memory volumes with multi-modal content and optimizing the trade-off between performance, retrieval depth, latency, and token efficiency.